\documentclass[journal]{IEEEtran}

\usepackage{xcolor,soul,framed} 

\colorlet{shadecolor}{yellow}
\usepackage[pdftex]{graphicx}
\graphicspath{{../pdf/}{../jpeg/}}
\DeclareGraphicsExtensions{.pdf,.jpeg,.png}

\usepackage[cmex10]{amsmath}
\usepackage{array}
\usepackage{mdwmath}
\usepackage{mdwtab}
\usepackage{multirow}
\usepackage{subcaption}
\usepackage{eqparbox}
\usepackage{url}
\usepackage{makecell}
\begin{document}
\bstctlcite{IEEEexample:BSTcontrol}
    \title{Automatic Diagnosis of Schizophrenia in EEG Signals Using CNN-LSTM Models}
  \author{Afshin Shoeibi,~\IEEEmembership{}
      Delaram Sadeghi,~\IEEEmembership{}
      Parisa Moridian,~\IEEEmembership{}
      Navid Ghassemi,~\IEEEmembership{}
      Jonathan Heras,~\IEEEmembership{}
      Roohallah Alizadehsani,~\IEEEmembership{}
      Ali Khadem,~\IEEEmembership{}
      Yinan Kong,~\IEEEmembership{}
      Saeid Nahavandi,~\IEEEmembership{}
      Yu-Dong Zhang,~\IEEEmembership{}
      and~Juan M. Gorriz ~\IEEEmembership{}

  \thanks{ A. Shoeibi, N. Ghassemi and A. Khadem are with the Faculty of Electrical Engineering, K. N. Toosi University of Technology, Tehran, Iran. (e-mail: afshin.shoeibi@gmail.com).}
  \thanks{ D. Sadeghi is with the Dept. of Medical Engineering, Mashhad Branch, Islamic Azad University, Mashhad, Iran.}%
  \thanks{ P. Moridian is with the Faculty of Engineering, Science and Research Branch, Islamic Azad University, Tehran, Iran.}%
  \thanks{ J. Heras is with the Department of Mathematics and Computer Science, University of La Rioja, La Rioja, Spain.}%
  \thanks{ R. Alizadehsani is with the Institutefor Intelligent Systems Research and Innovation (IISRI), Deakin University,Victoria 3217, Australia.}%
  \thanks{ Y. Kong is with the School of Engineering, Macquarie University, Sydney 2109, Australia.}%
  \thanks{ S. Nahavandi. is with the Institute for Intelligent Systems Research and Innovation (IISRI), Deakin University,Victoria 3217, Australia. Also with the Harvard Paulson School of Engineering and Applied Sciences, Harvard University, Allston, MA 02134 USA.}%
  \thanks{ Yu-Dong Zhang is with the Department of Informatics, University of Leicester, Leicester, United Kingdom.}%
  \thanks{ Juan M. Gorriz is with the Department of Signal Theory, Networking and Communications, Universidad de Granada, Spain. Also with the Department of Psychiatry. University of Cambridge, UK.}}%


\maketitle

\begin{abstract}
Schizophrenia (SZ) is a mental disorder whereby due to the secretion of specific chemicals in the brain, the function of some brain regions is out of balance, leading to the lack of coordination between thoughts, actions, and emotions. This study provides various intelligent deep learning (DL)-based methods for automated SZ diagnosis via electroencephalography (EEG) signals. The obtained results are compared with those of conventional intelligent methods. To implement the proposed methods, the dataset of the Institute of Psychiatry and Neurology in Warsaw, Poland, has been used. First, EEG signals were divided into 25 s time frames and then were normalized by z-score or norm L2. In the classification step, two different approaches were considered for SZ diagnosis via EEG signals. In this step, the classification of EEG signals was first carried out by conventional machine learning methods, e.g., support vector machine, k-nearest neighbors, decision tree, na\"ive Bayes, random forest, extremely randomized trees, and bagging. Various proposed DL models, namely, long short-term memories (LSTMs), one-dimensional convolutional networks (1D-CNNs), and 1D-CNN-LSTMs, were used in the following. In this step, the DL models were implemented and compared with different activation functions. Among the proposed DL models, the CNN-LSTM architecture has had the best performance. In this architecture, the ReLU activation function with the z-score and L2-combined normalization was used. The proposed CNN-LSTM model has achieved an accuracy percentage of 99.25
\end{abstract}



%
 \IEEEpeerreviewmaketitle


\section{Introduction}

\IEEEPARstart{S}{Schizophrenia} (SZ) is one of the most important mental disorders, leading to disruption in brain growth \cite{a1,a2}. This disorder seriously damages thoughts \cite{a3}. The reason for SZ is not fully understood, though most research has demonstrated that the brain's structural and functional abnormalities play a role in its creation \cite{a4}. According to the World Health Organization (WHO) reports, nearly 21 million individuals suffer from such a brain disorder worldwide. The average age starting to get affected by this disorder is in youth age; in males 18 years old, and females 25 years old, and it is more prevalent among males \cite{a5}.

Numerous methods have been provided for automated SZ diagnosis; among these techniques, neuroimaging-based methods have a special potential for specialist physicians \cite{a6,a7}. Generally, neuroimaging methods include various structural or functional modalities \cite{a8,a9}. Structural Magnetic Resonance Imaging (sMRI) and DTI-MRI are among the most important modalities of structural neuroimaging, providing important information regarding brain structure to specialist physicians \cite{a10,a11,a12}. Contrarily, EEG \cite{a13}, magnetoencephalography (MEG) \cite{a14}, functional MRI (fMRI) \cite{a15}, and functional near-infrared spectroscopy (fNIRS) \cite{a16} are the most important functional modalities of the brain. These modalities provide vital information on brain function to specialist physicians. 

\begin{figure*}[ht]
\centering
\includegraphics[width=7in]{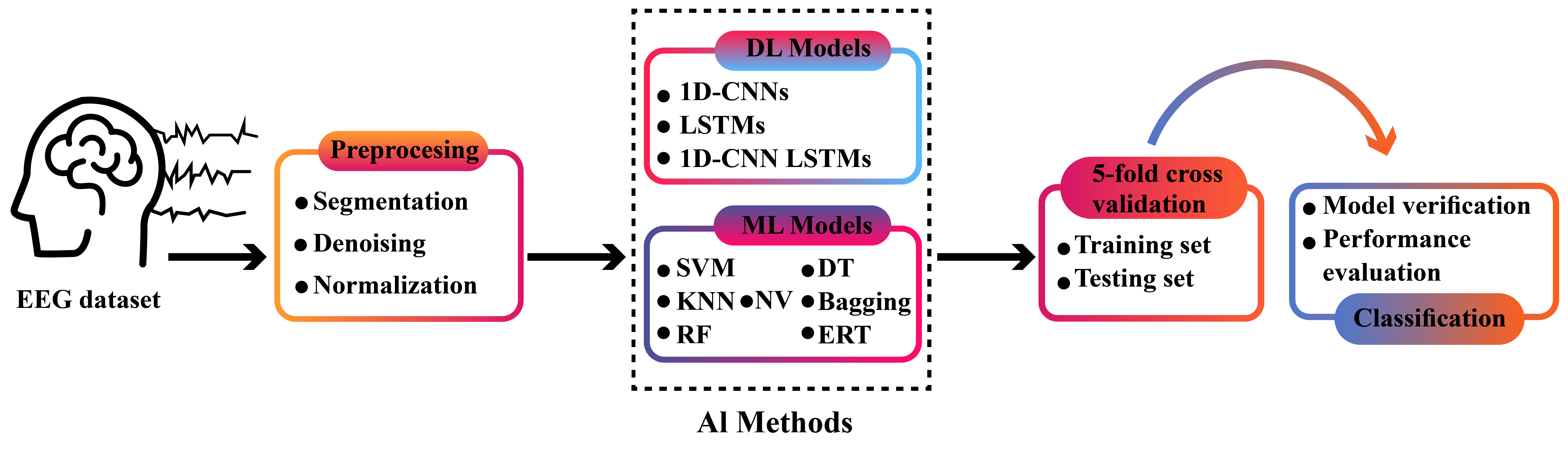}
\caption{The block diagram of proposed methods.}
\label{fig:1}
\end{figure*}

 EEG is one of the most practical and inexpensive functional neuroimaging modalities, specifically capturing specialist physicians' interests. In this modality, the brain's electrical activities are recorded from the head surface with a high temporal resolution and an appropriate spatial resolution, which is influential in SZ diagnosis \cite{a17}. In addition to the mentioned merits, EEG signals regularly have various channels recorded in long-term \cite{a17}. In some cases, these reasons make specialist physicians face serious challenges in SZ diagnosis via EEG signals. 

In recent years, various investigations have provided automated SZ diagnosis via EEG signals using Artificial Intelligence (AI) methods \cite{a18,a19,a20,a21}. The AI investigations in this field include conventional machine learning and DL methods \cite{a18,a19,a20,a21}. The AI-based SZ diagnosis algorithm includes preprocessing sections, features extraction and selection, and in the end, classification. Feature extraction is the most important part of SZ diagnosis via EEG signals. In conventional machine learning, the extracted features from EEG signals are mainly categorized into four groups: time \cite{a22}, frequency \cite{a23}, time- frequency \cite{a24}, and non-linear \cite{a25,a26} fields. Siuly et al. \cite{a27} used empirical mode decomposition (EMD) in preprocessing step. In the following, various statistical features were extracted from EMD sub-bands, and the EBT method was used for classification. In another study, Jahmunaha et al. \cite{a28} used non-linear features and SVM with radial basis function (RBF) kernel (SVM-RBF) in the feature extraction and classification steps, respectively. Devia et al. \cite{a29} have provided an Event related field (ERF) features-based SZ diagnosis method via EEG signals. ERT features were extracted from EEG signals in this effort, and then linear discriminant analysis (LDA) was used in the classification step. In \cite{a30}, Statistical features of steady-state visual evoked potentials (SSVEPs) were extracted, and in the end, classification has been executed by the KNN method. Fali et al. \cite{a32} used SPN features and SVM classification for SZ diagnosis via EEG signals. In another study, Shim et al. provided a new method of SZ diagnosis via EEG signals \cite{a35}. This investigation used sensor-level and source-level features in the feature extraction step and then employed the Fisher score for feature selection. Ultimately, the SVM method was used in the classification step, and they achieved promising results. 

In conventional machine learning, selecting proper feature extraction algorithms for SZ diagnosis is a relatively demanding task, requiring a great deal of knowledge in signal processing and the AI field. In order to overcome this problem, DL-based methods have been provided in recent years for SZ diagnosis via EEG signals, where feature extraction operations are carried out without surveillance by deep layers \cite{a25}. Shalbaf et al. \cite{a36} define a transfer learning model for SZ diagnosis via EEG signals. In this study, the ResNet-18 model has been used for feature extraction from EEG signals. Besides, SVM has been used in the classification step. Some researchers have studied other CNN models utilization in SZ diagnosis via EEG signals. CNN models have been used in \cite{a41,a42} for SZ diagnosis, resulting in satisfactory achievements. Convolutional recurrent neural network (CNN-RNN) models are an important group of DL networks and are significantly popular for their capability of various brain diseases diagnoses via EEG signals. In \cite{a37,a38,a43,a44}, CNN-LSTM models have been used for SZ diagnosis, and the researchers have been able to achieve promising results.

In this paper, SZ diagnosis via EEG signals will be investigated by using various proposed DL and conventional ML-based methods. A summary of proposed methods is depicted in \ref{fig:1}. In this study, the dataset of the Institute of Psychiatry and Neurology in Warsaw, Poland, is used \cite{a46}. In the preprocessing step, the z-score and L2 normalization techniques will be applied to EEG signals. Next, in order to classify EEG signals, various conventional ML methods and DL-based proposed models will be used. The conventional ML methods employed, include various classification, SVM \cite{a47}, KNN \cite{a48}, DT \cite{a49}, naïve Bayes \cite{a50}[50], RF \cite{a51}, ERT \cite{a52}, and bagging \cite{a53} methods. Besides, the proposed DL networks include various 1D-CNN, LSTM, ID-CNN-LSTM models for executing the steps from feature extraction to classification. Generally, 9 LSTM, 1D-CNN, and ID-CNN-LSTM-based DL methods will be investigated in this step. 

In section II, we described our method in detail. In addition, we outline several baseline methods for comparison purposes in the same section. The statistical metrics to analyze and validate the proposed model are described in section III. Experiment Results are provided in section IV, and some limitations of the proposed method are provided in section V. Finally, a discussion, the conclusion, and future works are represented in section VI.
\section{Material and Methods}
This section will discuss the proposed methods for SZ diagnosis via EEG signals and various conventional ML and DL models. First, the proposed dataset will be examined. Then, the preprocessing method of EEG signals will be explained. In the end, conventional ML and DL models will be introduced for SZ diagnosis via EEG signals. 

\subsection {Dataset}
\begin{figure}[t]
\centering
\includegraphics[width=3.5in]{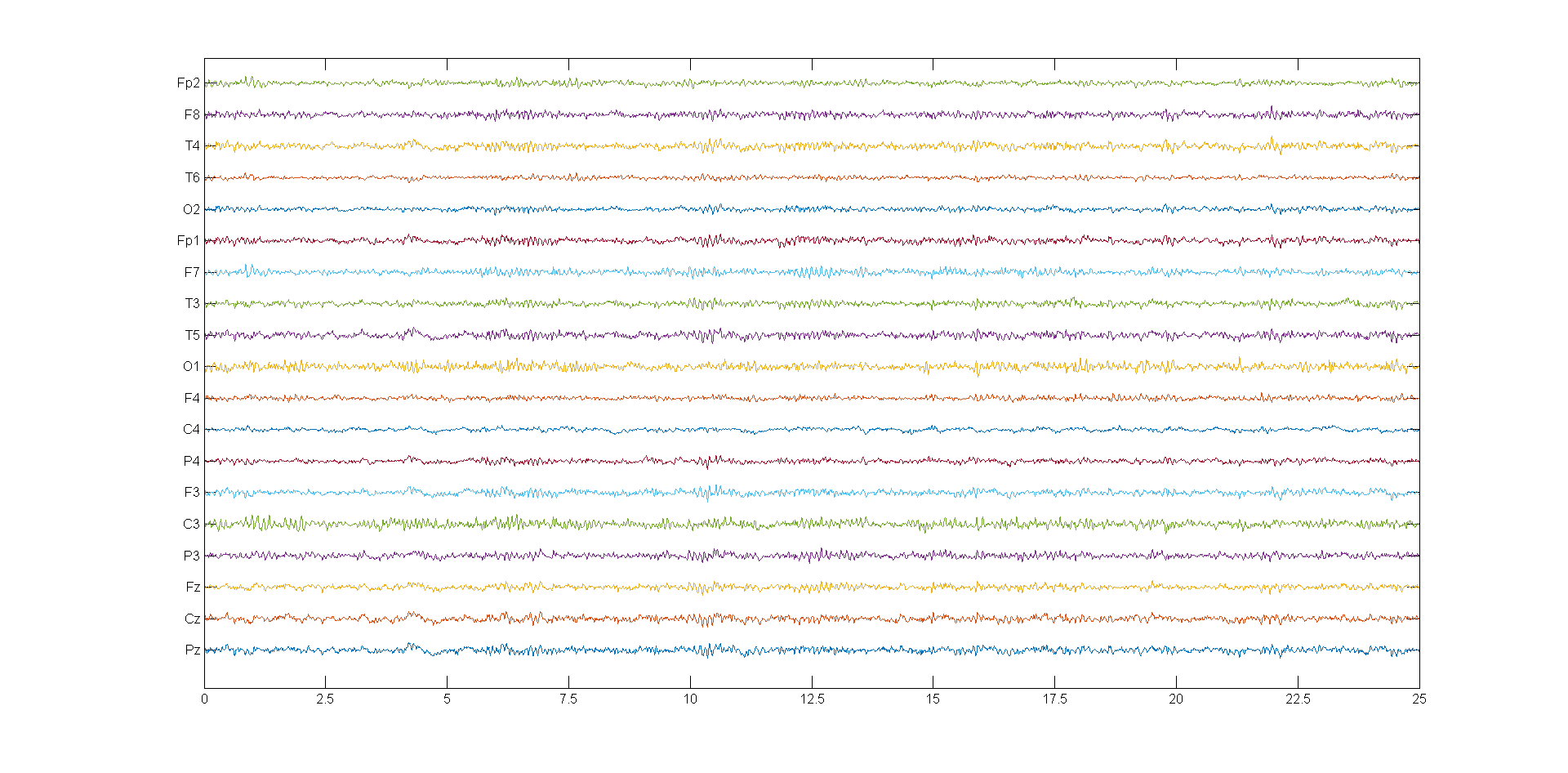}
\caption{A sample frame of the EEG signals of a person with SZ.}
\label{fig:2}
\end{figure}

\begin{figure}[t]
\centering
\includegraphics[width=3.5in]{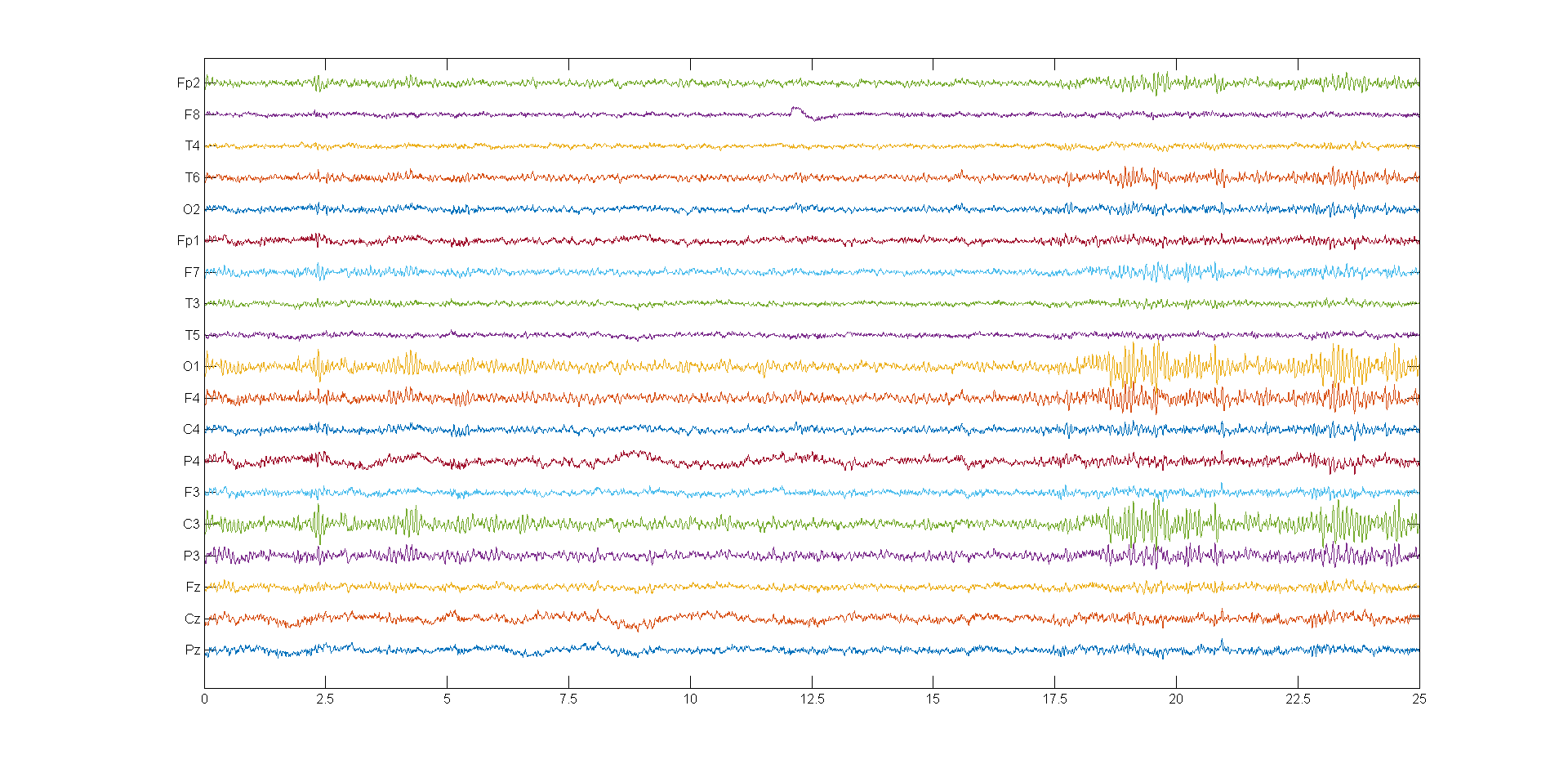}
\caption{A sample frame of the EEG signals of a normal person.}
\label{fig:3}
\end{figure}

This dataset includes recorded EEG signals from 14 females and males with ages between 27.9 and 28.3 yrs. Besides, 14 normal individuals matched with the patients in terms of age and gender were employed in this institution, and the data recording was carried out \cite{a46}. A signal recording was performed with the eyes closed in 15 minutes for each case. Recording EEG signals was performed by using standard 10-20 with a sampling frequency of 250 Hz \cite{a46}. In this study, the used electrodes include Fp1, Fp2, F7, F3, Fz, F4, F8, T3, C3, Cz, C4, T4, T5, P3, Pz, P4, T6, O1, and O2. An example of EEG signals of SZ and normal cases is depicted in Fig. \ref{fig:2} and \ref{fig:3}. 

\subsection {Preprocessing}

In order to pre-process the EEG signals of the mentioned dataset, several steps are used. First, each 19 recorded EEG signal has been divided into overlap-free 25-seconds frames, each of which includes 6250 temporal samples. Accordingly, each frame of EEG signals has 6250×19 dimensions. In the following, each EEG frame has been normalized by z-score and L2 methods. The normalization of EEG signals helps the accuracy and performance enhancement in conventional ML and DL models. 

\subsection {Conventional Machine Learning Methods}

The proposed conventional ML methods are introduced in this section as a baseline for comparison purposes. The proposed algorithms include SVM \cite{a47}, KNN \cite{a48}, DT \cite{a49}, naïve Bayes \cite{a50}, RF \cite{a51}, ERT \cite{a52}, and bagging \cite{a53}. Each of these methods will be briefly introduced in the following. 

\subsubsection {Support Vector Machine}
SVM \cite{a47} is an algorithm that constructs a hyper-plane or set of hyper-planes in a high or infinite dimensional space, which can be used for classification, regression or other tasks. Intuitively, a good separation is achieved by the hyper-plane that has the largest distance to the nearest training data points of any class (so-called functional margin), since in general the larger the margin the lower the generalization error of the classifier.

\subsubsection {K-Nearest Neighbors}

KNN \cite{a48} is a classification algorithm where some fixed and small number (k) of nearest neighbors (based on a notion of distance) from the training set are located and used together to determine the class of the test instance through a simple majority voting; that is, the class of the test instance is assigned the data class which has the most representatives within the k nearest neighbors of that point. 
\subsubsection {Decision Tress}

DT \cite{a49} is an algorithm that creates a model that predicts the class of an instance by learning simple decision rules inferred from the data features. The representation of a DT model is a binary tree wherein each node represents a single input variable (X) and a split point on that variable, assuming the variable is numeric. The leaf nodes (also called terminal nodes) of the tree contain an output variable (y) which is used to make a prediction.

\subsubsection {Naïve Bayes}

Naive Bayes \cite{a50} is a supervised learning algorithm based on applying Bayes’ theorem with the “naive” assumption of conditional independence between every pair of features given the value of the class variable. This means that we calculate P(data|class) for each input variable separately and multiple the results together, for example: P(class | X1, X2, …, Xn) = P(X1|class) × P(X2|class) × … × P(Xn|class) × P(class) / P(data); where P(A | B) represents the probability of A given B. 

\subsubsection {Random Forest}

Random Forest (RF) \cite{a51} is an extension of the bagging algorithm where a number of DT classifiers are fit on various sub-samples of the dataset and uses averaging to improve the predictive accuracy and control over-fitting. Unlike bagging, random forest also involves selecting a subset of input features (columns or variables) at each split point in the construction of trees. By reducing the features to a random subset that may be considered at each split point, it forces each decision tree in the ensemble to be more different.

\subsubsection {Extremely Randomized Trees}

Extremely Randomized Trees (ERT) \cite{a52}, as RF, is an ensemble of several DT models. However, the ERT algorithm fits each decision tree on the whole training dataset instead of using a bootstrap sample. Like the RF algorithm, the ERT algorithm will randomly sample the features at each split point of a decision tree; but instead of using a greedy algorithm to select an optimal split point, the ERT selects a split point at random.

\subsubsection {Bagging}

Bagging \cite{a53} is an ensemble classifier that fits base classifiers on random subsets of the original dataset and then aggregate their individual predictions (either by voting or by averaging) to form a final prediction. To be more concrete, in bagging, a number of classifiers are created where each classifier is created from a different bootstrap sample of the training dataset. A bootstrap sample is a sample of the training dataset where a sample may appear more than once in the sample, referred to as sampling with replacement.

\subsection {Deep Learning Models}

This section provides various types of 1D-CNN, LSTM, and 1D-CNN-LSTM models for SZ diagnosis via EEG signals. Various types of the suggested 1D-CNN, LSTM, and 1D-CNN-LSTM models will be examined in the following. 

\subsubsection {ID-CNN models}

The higher performance of CNN models in machine vision has led them to be used in time series processing, such as medical signals, leading to successful results \cite{a54,a55}. The CNN models have important convolutional, pooling, and fully connected (FC) layers \cite{a56,a57}. In 1D-CNN models, signal time can be considered a spatial dimension, e.g., height or width of a 2D image \cite{a58}. 1D-CNN models are considered the important rivals of RNN architectures in time series processing. Compared to RNN models, 1D-CNN architectures have lower computational costs \cite{a58}. In this section, the three proposed 1D-CNN-based models are provided for SZ diagnosis via EEG signals.

\textbf{(A) The first version of 1D-CNN model}

The details of the first proposed 1D-CNN model are provided in Table \ref{tab:1}. Concerning Table \ref{tab:1}, this model includes nine different layers. The convolutional layers have 64 filters with 3×3 dimensions. In addition, various activation functions, e.g., ReLU, Leaky ReLU, and seLU, have been used in convolutional layers, and the related results will be compared in the Experiment Results section. Besides, a max-pooling layer has been used for decreasing dimensions, dropout layers with different rates for the prevention of overfitting, flatten layer for converting a matrix to vector, and in the end, dense layers for classification. The activation function of the final dense layer is of sigmoid type, used for binary classification. 
\begin{table}[]
\centering
\caption{The details of the first proposed 1D-CNN model}
\begin{tabular}{lcccc}

\hline
Layers      & Filters & Kernel Size & Stride           & Activation \\ 
\hline
Input Data  & --     & --         &--               &-- \\
Conv1D      & 64      & 3           & 1                & ReLU       \\
Conv1D      & 64      & 3           & 1                & ReLU       \\
Dropout     & --     & --         & Rate=0.25        & --            \\
Max Pooling & --     & 2           & 1                & --          \\
Flatten     & --     & --         & --              & --          \\
Dense       & 100     & --         & --              & --          \\
Dropout     & --     & --         & Rate=0.25        & --          \\
Dense       & 2       & --         & --              & Sigmoid    \\ 
\hline
\end{tabular}
\label{tab:1}
\end{table}

\textbf{(B) The second version of 1D-CNN model}

The architecture of the second proposed 1D-CNN model has three convolutional layers, and their filters' number, kernel size, and activation function have been indicated in Table \ref{tab:2}. In this model, a convolutional layer with a kernel size of 2 has been used. Moreover, this model has four dropout layers with different rates, 1 flatten layer and two dense layers. The activation function of the first dense layer is of ReLU type, and the activation function of the final dense layer is for sigmoid classification. 


\begin{table}[]
\centering
\caption{The details of the second proposed 1D-CNN model}
\begin{tabular}{lcccc}

\hline
Layers      & Filters    & Kernel Size    & Stride        & Activation \\ 
\hline
Input Data  & --         & --            &--           &-- \\
Conv1D      & 64          & 3              & 1            & ReLU       \\
Dropout     & --         & --            & Rate=0.5     & --       \\
Conv1D      & 64          & 3              & 1            & ReLU      \\
Dropout     & --         & --            & Rate=0.5     & --          \\
Conv1D      & 64          & 3              & 1            & ReLU          \\
Dropout      & --        & --            & Rate=0.5     & --          \\
Max Pooling     & --     & 2              & 1            & --          \\
Flatten     & --         & --            & --          & --          \\
Dense       & 100         & --            & --          & ReLU          \\
Dropout     & --         & --            & Rate=0.25    & --          \\
Dense       & 1           & --            & --          & Sigmoid    \\ 
\hline
\end{tabular}
\label{tab:2}
\end{table}

\textbf{(C) The Third version of 1D-CNN model}

According to Table \ref{tab:3}, the third proposed 1D-CNN model consists of two convolutional layers with a similar number of filters, kernel size, and activation functions to the previous networks. This model has a Max Pooling layer with a kernel size of 2. In addition, it takes advantage of dropout with different rates.  Similar to previous models, a flatten layer is also used in this model. This model consists of two dense layers, in which the activation functions of the first and second layers are of ReLU and sigmoid type, respectively.

\begin{table}[]
\centering
\caption{The details of the third proposed 1D-CNN model}
\begin{tabular}{lcccc}

\hline
Layers      & Filters & Kernel Size & Stride           & Activation \\ 
\hline
Input Data  & --     & --         &--               &-- \\
Conv1D      & 64      & 3           & 1                & ReLU       \\
Conv1D      & 64      & 3           & 1                & ReLU       \\
Dropout     & --     & --         & Rate=0.5         & --            \\
Max Pooling & --     & 2           & 1                & --          \\
Flatten     & --     & --         & --              & --          \\
Dense       & 100     & --         & --              & ReLU          \\
Dropout     & --     & --         & Rate=0.25        & --          \\
Dense       & 50      & --         & --              & ReLU          \\
Dropout     & --     & --         & Rate=0.25        & --          \\
Dense       & 1       & --         & --              & Sigmoid    \\ 
\hline
\end{tabular}
\label{tab:3}
\end{table}

\subsubsection{LSTM models}

RNNs are a group of DL models employed in speech recognition \cite{a59}, natural language processing (NLP) \cite{a60}, and biomedical signal processing \cite{a61,a62}. CNN models are of Feed-Forward types. However, the RNNs have a FeedBack layer, in which the network output returns to the network along with the next input. Because of having internal memory, RNNs memorize their previous input and use it to process a sequence of inputs. Simple RNN, LSTM, and GRU networks are three important groups of RNNs \cite{a58}. In this section, various LSTM models of SZ diagnosis via EEG signals will be proposed.   

\textbf{(A) The first version of LSTM model}

In Table \ref{tab:4}, the details of the first proposed LSTM model consisting of 6 layers are presented. In this model, an LSTM layer with a kernel size of 100 is employed.  Another section of the proposed LSTM architecture consists of two different layers of dropout and rate and two dense layers. In the first and second dense layers, the ReLU and sigmoid activation functions are used.

\begin{table}[]
\centering
\caption{The details of the first proposed LSTM model}
\begin{tabular}{lcccc}

\hline
Layers      & Filters & Kernel Size & Stride           & Activation \\ 
\hline
Input Data  & --     & --         &--               &-- \\
LSTM        & 100    & 1          & --              & --       \\
Dropout     & --     & --         & Rate=0.5       & --            \\
Dense       & 100     & --         & --              & ReLU          \\
Dropout     & --     & --         & Rate=0.25        & --          \\
Dense       & 1       & --         & --              & Sigmoid    \\ 
\hline
\end{tabular}
\label{tab:4}
\end{table}

\textbf{(B) The second version of LSTM model}

In Table \ref{tab:5}, the details of the second proposed LSTM model consisting of 7 layers are presented. In this architecture, an LSTM layer with a kernel size of 50 is added to the previous model. The reason behind this is to examine the effect of adding LSTM layers on SZ diagnosis accuracy via EEG signals. 

\begin{table}[]
\centering
\caption{The details of the second proposed LSTM model}
\begin{tabular}{lcccc}

\hline
Layers      & Filters & Kernel Size & Stride           & Activation \\ 
\hline
Input Data  & --     & --         &--               &-- \\
LSTM        & 100    & 1          & --              & --       \\
LSTM        & 50    & 1          & --              & --       \\
Dropout     & --     & --         & Rate=0.5       & --            \\
Dense       & 100     & --         & --              & ReLU          \\
Dropout     & --     & --         & Rate=0.25        & --          \\
Dense       & 1       & --         & --              & Sigmoid    \\ 
\hline
\end{tabular}
\label{tab:5}
\end{table}

\subsubsection{CNN-LSTM models}

In CNN-RNN models, the convolutional layers are used in the first layers of the model to extract the features and find the local patterns \cite{a58}. Then, their outputs are applied to RNN layers. Experimentally, the convolutional layers extract the local and spatial patterns of EEG signals better compared to RNNs. Besides, adding convolutional layers to RNN allows a more accurate examination of data. In this section, various CNN-LSTM models for SZ diagnosis will be proposed. 

\textbf{(A) The first version of CNN-LSTM model}

The first proposed CNN-LSTM model consists of 11 Max, dropout, CNN, LSTM, flatten, pooling, and dense layers. The details of the proposed model are presented in Table \ref{tab:6}. This architecture includes two convolutional layers; three dropout layers with different rates, one Max-Pooling layer, and one flatten layer, one LSTM layer, and finally, two dense layers with ReLU and sigmoid activation functions. 

\begin{table}[]
\centering
\caption{The details of the first proposed CNN-LSTM model}
\begin{tabular}{lcccc}

\hline
Layers      & Filters & Kernel Size & Stride           & Activation \\ 
\hline
Input Data  & --     & --         &--               &-- \\
Conv1D      & 64      & 3           & 1                & ReLU       \\
Conv1D      & 64      & 3           & 1                & ReLU       \\
Dropout     & --     &             & Rate=0.5        & --            \\
Max Pooling & --     & 2           & 1                & --          \\
Flatten     & --     & --         & --              & --          \\
LSTM        & 100     & 1           & --              & --          \\
Dropout     & --     &             & Rate=0.5        & --            \\
Dense       & 100     & --         & --              & --          \\
Dropout     & --     & --         & Rate=0.25        & --          \\
Dense       & 1       & --         & --              & Sigmoid    \\ 
\hline
\end{tabular}
\label{tab:6}
\end{table}

\textbf{(B) The second version of CNN-LSTM model}

\begin{figure}[ht]
\centering
\includegraphics[width=3.5in]{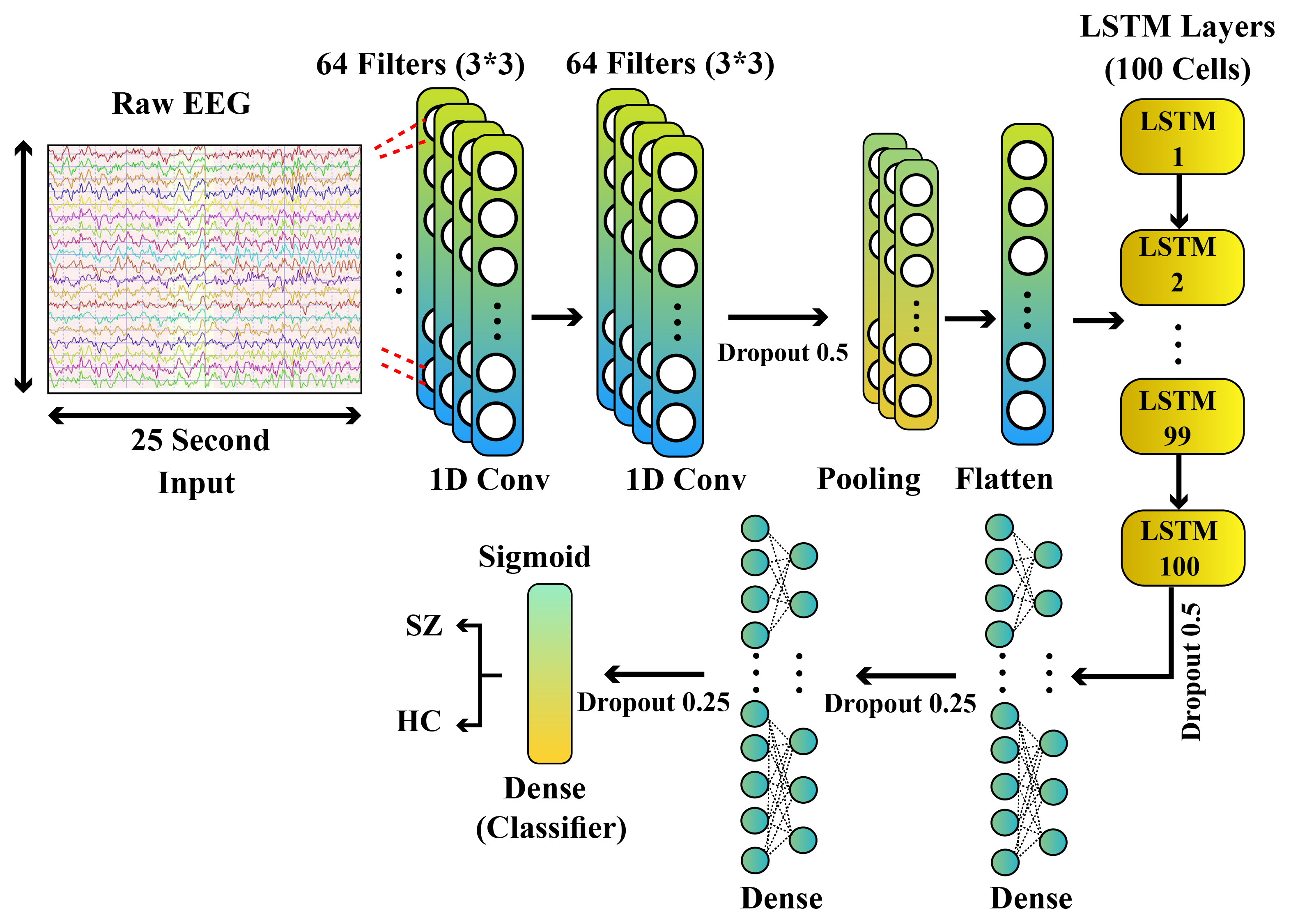}
\caption{The second version of the proposed CNN-LSTM model for diagnosis of SZ.}
\label{fig:4}
\end{figure}

In this section, the second proposed CNN-LSTM model will be introduced. This network includes 13 layers, and similar to the previous model, it consists of CNN and LSTM layers whose details are demonstrated in Table \ref{tab:7} and Figure \ref{fig:4}. As can be seen in Table \ref{tab:7} and Figure \ref{fig:4}, the first ten layers of this proposed model are identical to those of the previous CNN-LSTM model. The dense layer with 50 neurons and the ReLU activation function are used in the 11th layer of this architecture. The 12th layer comprises dropout with rate=0.25. Ultimately, in the 13th layer, the dense layer with a sigmoid activation function for classification is employed. 

\begin{table}[]
\centering
\caption{The details of the second proposed CNN-LSTM model}
\begin{tabular}{lcccc}

\hline
Layers      & Filters & Kernel Size & Stride           & Activation \\ 
\hline
Input Data  & --     & --         &--               &-- \\
Conv1D      & 64      & 3           & 1                & ReLU       \\
Conv1D      & 64      & 3           & 1                & ReLU       \\
Dropout     & --     &             & Rate=0.5        & --            \\
Max Pooling & --     & 2           & 1                & --          \\
Flatten     & --     & --         & --              & --          \\
LSTM        & 100     & 1           & --              & --          \\
Dropout     & --     &             & Rate=0.5        & --            \\
Dense       & 100     & --         & --              & --          \\
Dropout     & --     & --         & Rate=0.25        & --          \\
Dense       & 50       & --         & --              & ReLU    \\ 
Dropout     & --     & --         & Rate=0.25        & --          \\
Dense       & 1       & --         & --              & Sigmoid    \\ 
\hline
\end{tabular}
\label{tab:7}
\end{table}
\begin{table}[!h]
    \centering
    \caption{Description of statistical parameters used.}
    \begin{tabular}{|c|c|}
        \hline
        Parameters Name & Formula \\
        \hline
        Accuracy & Acc = $\frac{\text{TP}+\text{TN}}{\text{FP}+\text{FN}+\text{TP}+\text{TN}} $ \\
        \hline
        Precision &  Prec = $\frac{\text{TP}}{\text{TP}+\text{FP}} $ \\
        \hline
        Recall &  Rec = $\frac{\text{TP}}{\text{TP}+\text{FN}} $ \\
        \hline
    \end{tabular}
    
    \label{tab:metrics}
\end{table}
\section{Statistical Metrics}
In this work, a 5-fold cross-validation method is used to obtain the results; the advantage of K-fold cross-validation is that all the data points of the dataset are used for both training and testing, and also the results are more reliable. The performance of each algorithm is evaluated using three different statistical metrics, namely, accuracy (Acc), Precision (Prec), and recall (Rec), all of which are shown in Table \ref{tab:metrics}. These metrics are extracted from the confusion matrix: true positive (TP), false negative (FN), true negative (TN), and false positive (FP) \cite{a25}. Additionally, the area under the ROC curve (AUC) is calculated for each method as well.
\section{Experiment Results}
\begin{table*}[ht]
\centering
\caption{Performance criteria of the proposed ML methods.}
\resizebox{\linewidth}{!}{
\begin{tabular}{|c|c|c|c|c|c|c|c|c|}
\hline
\multicolumn{1}{|c|}{\multirow{2}{*}{Methods}} & \multicolumn{4}{c|}{Raw EEG} & \multicolumn{4}{c|}{Z-Score Normalized EEG} \\
\cline{2-9}
\multicolumn{1}{|c|}{} & \multicolumn{1}{c|}{Acc} & \multicolumn{1}{c|}{Prec} & \multicolumn{1}{c|}{Rec} & \multicolumn{1}{c|}{AUC} & \multicolumn{1}{c|}{Acc} & \multicolumn{1}{c|}{Prec} & \multicolumn{1}{c|}{Rec} & \multicolumn{1}{c|}{AUC} \\
\hline
KNN     & 57.03 ± 2.21 & 52.12 ± 2.66  & 99.80 ± 0.38  & 59.58 ± 0.56 & 55.10 ± 1.77          & 49.32 ± 1.42 & 99.80 ± 0.39 & 60.13 ± 1.28 \\ \hline
DT      & 64.19 ± 3.08 & 62.49 ± 5.15  & 59.52 ± 5.40  & 63.94 ± 3.12 & 64.71 ± 4.12          & 59.28 ± 5.00 & 61.16 ± 5.14 & 64.31 ± 4.21 \\ \hline
SVM     & 54.14 ± 3.97 & 20.77 ± 25.50 & 32.57 ± 39.96 & 54.10 ± 5.16 & 62.09 ± 2.75          & 54.72 ± 2.92 & 77.81 ± 2.01 & 63.89 ± 2.42 \\ \hline
Bayes   & 62.62 ± 2.52 & 56.08 ± 2.76  & 93.21 ± 4.60  & 64.35 ± 2.30 & 59.12 ± 3.26          & 51.78 ± 2.38 & 94.81 ± 2.61 & 63.15 ± 2.97 \\ \hline
Bagging & 77.37 ± 3.23 & 81.80 ± 2.56  & 66.93 ± 6.13  & 76.91 ± 2.96 & \textbf{81.22 ± 1.74} & 82.90 ± 3.76 & 72.02 ± 1.95 & 80.21 ± 1.65 \\ \hline
RF      & 75.19 ± 2.19 & 83.60 ± 4.22  & 59.00 ± 3.62  & 74.20 ± 1.43 & 78.77 ± 1.55          & 81.23 ± 2.31 & 66.80 ± 2.94 & 77.44 ± 1.74 \\ \hline
ET      & 76.24 ± 1.84 & 80.64 ± 3.37  & 64.96 ± 2.10  & 75.57 ± 1.52 & 76.94 ± 1.81          & 76.29 ± 2.27 & 68.35 ± 3.90 & 75.96 ± 2.05 \\ \hline

\end{tabular}}
\label{tab:ml}
\end{table*}

\begin{figure*}[ht]
\centering
\begin{subfigure}[b]{0.39\textwidth}
    \centering
    \includegraphics[width=\textwidth]{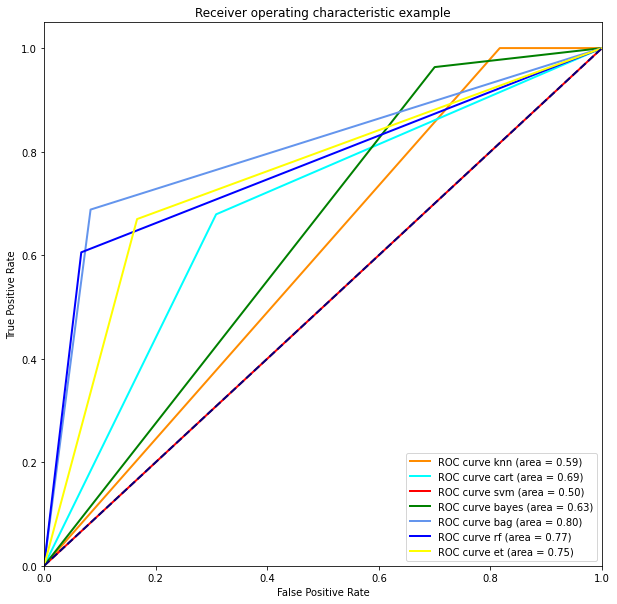}
    \caption{}
    \label{fig:5a}
\end{subfigure}
\hspace{7pt}
\begin{subfigure}[b]{0.39\textwidth}
    \centering
    \includegraphics[width=\textwidth]{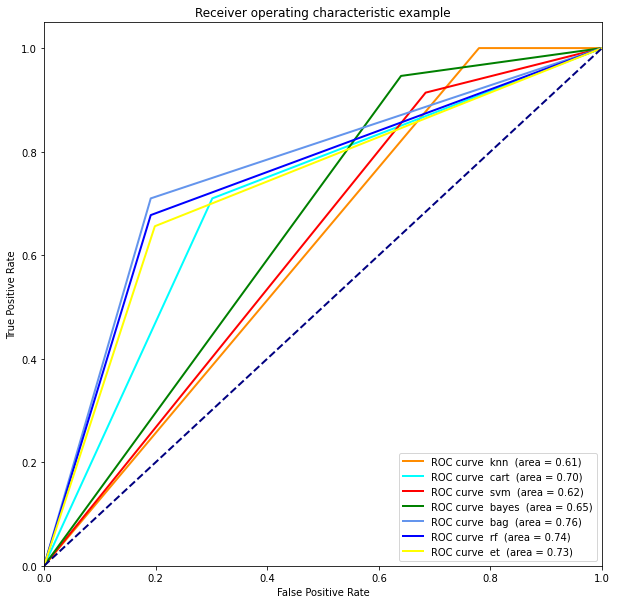}
    \caption{}
    \label{fig:5b}
\end{subfigure}
\caption{ROC curves of conventional ML classifiers}
\label{fig:5}
\end{figure*}

\begin{table}[t]
\centering
\caption{The final selected values for batch size and hyper-parameters of the proposed DL networks.}
\begin{tabular}{|c|c|c|c|}
\hline
Networks   & Epochs & Batch size & Learning rate \\ \hline
CNN-1      & 32     & 10         & 0.01          \\ \hline
CNN-2      & 32     & 10         & 0.01          \\ \hline
CNN-3      & 32     & 10         & 0.01          \\ \hline
LSTM-1     & 30     & 16         & 0.01          \\ \hline
LSTM-2     & 30     & 16         & 0.01          \\ \hline
CNN-LSTM 1 & 50     & 128        & 0.01          \\ \hline
CNN-LSTM 2 & 50     & 128        & 0.01          \\ \hline
\end{tabular}
\label{tab:9}
\end{table}
The results of the proposed methods are presented in this section. First, the simulation results obtained from conventional ML techniques for SZ diagnosis via EEG signals are presented and discussed.  The original dataset was flattened to have only a vector per sample, and then we used the flattened dataset to train several classification algorithms using the scikit-learn library \cite{a63}. Namely, we studied the performance of KNNs, DTs, SVMs, naive Bayes; and three ensemble algorithms (bagging, extremely randomized trees, and random forest). The algorithms were trained using the by-default hyper-parameters provided by the implementation of the scikit-learn library. Moreover, we studied the impact of Z-score normalization \cite{a64} on the performance of the models. All the experiments were conducted in an Intel (R) Core (TM) i7-4810MQ CPU at 2.80GHz. In Table \ref{tab:ml}, the results obtained from conventional classification algorithms for raw input EEG signals or normalized by z-score normalization are indicated.  

 
According to Table \ref{tab:ml}, the bagging conventional classification algorithms for EEG signals normalized using z-score normalization resulted in the maximum accuracy.  Figures \ref{fig:5a} and \ref{fig:5b} demonstrate the receiver operating characteristic (ROC) diagrams of the conventional classification algorithms for the raw input EEG signals or normalized using z-score normalization.

We also employed several DL architectures based on CNNs and LSTMs \cite{a58}, and the combination of both convolutions and LSTM layers. Namely, 3 CNNs, 2 LSTMs and 2 CNN-LSTM networks (see Tables 1-7 for the concrete architecture of these networks) were studied. We also analyzed the relevance of using 3 different activation functions (ReLU, Leaky ReLU, and seLU), and the impact of Z-score normalization. In order to avoid overfitting, we applied two regularization techniques that are Dropout and weight regularization \cite{a58}. In particular, dropout was applied after each convolutional and LSTM layer using a dropout value of 0.5, and after dense layers using a dropout value of 0.25. Weight regularization was employed in all the convolutional, LSTM, and dense layers of our architectures using L2 regularization with value 0.01. The final selected values for batch size and hyper-parameters of our networks are all available in Table \ref{tab:9}. All the experiments were conducted using the Keras library \cite{a65} and using a GPU NVidia RTX2080 Ti.

In the following, the results obtained from the DL proposed methods for different activation functions are demonstrated in Table \ref{tab:ten} to \ref{tab:12}. First, the results obtained from the proposed DL method with the Leaky ReLU activation function are demonstrated in Table \ref{tab:ten}. 

\begin{table*}[ht]
\centering
\caption{Performance criteria of the proposed DL methods with Leaky ReLU activation function.}
\resizebox{\linewidth}{!}{
\begin{tabular}{|c|c|c|c|c|c|c|c|c|}

\hline
\multicolumn{1}{|c|}{\multirow{2}{*}{Methods}} & \multicolumn{4}{c|}{Leaky ReLU + Z-Score} & \multicolumn{4}{c|}{Leaky ReLU + Z-Score + L2} \\
\cline{2-9}
\multicolumn{1}{|c|}{} & \multicolumn{1}{c|}{Acc} & \multicolumn{1}{c|}{Prec} & \multicolumn{1}{c|}{Rec} & \multicolumn{1}{c|}{AUC} & \multicolumn{1}{c|}{Acc} & \multicolumn{1}{c|}{Prec} & \multicolumn{1}{c|}{Rec} & \multicolumn{1}{c|}{AUC} \\
\hline
CNN-1      & 70.83 ± 8.76 & 58.12 ± 8.23  & 98.86 ± 1.24  & 80.95 ± 8.72  & 64.10 ± 6.68          & 52.17 ± 4.72 & 99.31 ± 0.90  & 86.73 ± 9.86  \\ \hline
CNN-2      & 38.42 ± 0.00 & 38.42 ± 0.00  & 100.00 ± 0.00 & 50.00 ± 0.00  & 40.00 ± 1.76          & 39.03 ± 0.67 & 99.77 ± 0.45  & 52.21 ± 3.22  \\ \hline
CNN-3      & 56.85 ± 4.17 & 47.24 ± 2.57  & 99.54 ± 0.55  & 67.19 ± 5.60  & 58.07 ± 3.77          & 47.93 ± 2.24 & 100.00 ± 0.00 & 82.73 ± 9.98  \\ \hline
LSTM-1     & 83.32 ± 2.55 & 73.64 ± 3.41  & 88.63 ± 6.66  & 91.03 ± 2.02  & 72.31 ± 8.37          & 56.03 ± 29.3 & 51.59 ± 29.76 & 74.52 ± 12.28 \\ \hline
LSTM-2     & 79.91 ± 9.00 & 72.12 ± 11.82 & 85.68 ± 5.90  & 86.90 ± 8.10  & 76.68 ± 6.51          & 70.79 ± 9.95 & 76.82 ± 23.80 & 80.30 ± 9.38  \\ \hline
CNN-LSTM 1 & 74.06 ± 19.9 & 65.83 ± 27.45 & 58.40 ± 32.91 & 78.32 ± 20.99 & 94.76 ± 5.94          & 90.95 ± 10.6 & 98.86 ± 1.24  & 99.73 ± 0.21  \\ \hline
CNN-LSTM 2 & 79.04 ± 12.2 & 71.51 ± 25.93 & 58.40 ± 36.37 & 85.79 ± 16.62 & \textbf{97.73 ± 1.39} & 96.35 ± 3.55 & 97.95 ± 1.32  & 99.71 ± 0.15  \\ \hline
\end{tabular}}
\label{tab:ten}
\end{table*}

\begin{table*}[ht]
\centering
\caption{Performance criteria of the proposed DL methods with seLU activation function.}
\resizebox{\linewidth}{!}{
\begin{tabular}{|c|c|c|c|c|c|c|c|c|}
\hline
\multicolumn{1}{|c|}{\multirow{2}{*}{Methods}} & \multicolumn{4}{c|}{seLU + Z-Score} & \multicolumn{4}{c|}{seLU + Z-Score + L2} \\
\cline{2-9}
\multicolumn{1}{|c|}{} & \multicolumn{1}{c|}{Acc} & \multicolumn{1}{c|}{Prec} & \multicolumn{1}{c|}{Rec} & \multicolumn{1}{c|}{AUC} & \multicolumn{1}{c|}{Acc} & \multicolumn{1}{c|}{Prec} & \multicolumn{1}{c|}{Rec} & \multicolumn{1}{c|}{AUC} \\
\hline
CNN-1      & 61.65 ± 4.89          & 50.49 ± 3.98 & 95.90 ± 4.22  & 69.50 ± 4.06  & 65.67 ± 5.95          & 53.71 ± 5.05 & 94.31 ± 6.14  & 75.12 ± 5.90  \\ \hline
CNN-2      & 57.90 ± 2.48          & 32.43 ± 19.4 & 59.77 ± 46.99 & 56.51 ± 11.79 & 58.42 ± 5.43          & 38.38 ± 19.4 & 51.36 ± 44.52 & 58.17 ± 8.82  \\ \hline
CNN-3      & 62.09 ± 4.43          & 50.71 ± 3.11 & 93.18 ± 11.94 & 69.17 ± 3.67  & 66.46 ± 4.20          & 54.76 ± 4.37 & 88.18 ± 12.48 & 76.09 ± 4.23  \\ \hline
LSTM-1     & 74.84 ± 5.05          & 64.48 ± 5.57 & 77.50 ± 9.15  & 82.90 ± 5.55  & 70.13 ± 8.80          & 57.07 ± 16.6 & 58.86 ± 29.15 & 72.11 ± 13.72 \\ \hline
LSTM-2     & \textbf{83.58 ± 0.81} & 74.99 ± 1.81 & 86.13 ± 3.16  & 91.06 ± 0.52  & 79.65 ± 6.27          & 72.75 ± 10.1 & 79.31 ± 8.36  & 86.43 ± 5.56  \\ \hline
CNN-LSTM 1 & 59.73 ± 1.47          & 41.14 ± 6.92 & 8.40 ± 3.18   & 50.95 ± 2.38  & 58.42 ± 3.39          & 48.12 ± 2.10 & 99.73 ± 0.45  & 89.44 ± 1.57  \\ \hline
CNN-LSTM 2 & 59.65 ± 3.02          & 43.78 ± 8.60 & 10.90 ± 3.26  & 61.16 ± 5.17  & 57.64 ± 1.68 & 47.59 ± 1.00 & 100.00 ± 0.00 & 87.08 ± 3.84  \\ \hline
\end{tabular}}
\label{tab:eleven}
\end{table*}
\begin{figure*}[ht]
\centering
\begin{subfigure}[b]{0.39\textwidth}
    \centering
\includegraphics[width=\textwidth]{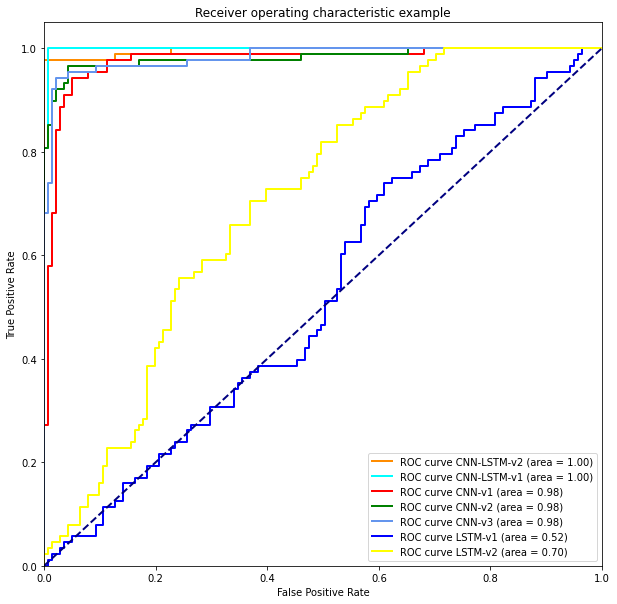}
    \caption{}
    \label{fig:6a}
\end{subfigure}
\hspace{7pt}
\begin{subfigure}[b]{0.39\textwidth}
    \centering
    \includegraphics[width=\textwidth]{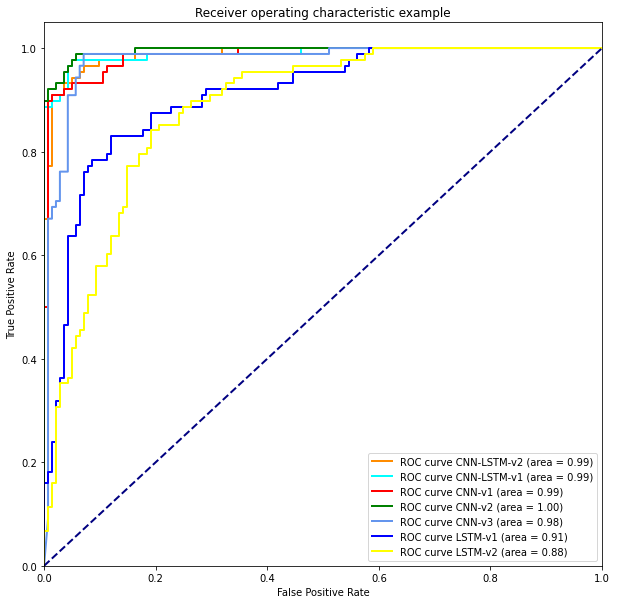}
    \caption{}
    \label{fig:6b}
\end{subfigure}
\caption{ROC curves of DL methods with ReLU activation function and z-score + L2 normalization}
\label{fig:6}
\end{figure*}
As indicated in Table \ref{tab:ten}, the second proposed CNN-LSTM model with the Leaky ReLU activation function and combined normalization of z-score with L2 could obtain the maximum accuracy. Table \ref{tab:eleven} presents the results obtained from the proposed DL method with the seLU activation function.

Table \ref{tab:eleven} indicated that the second proposed LSTM method could result in maximum accuracy. The results of all proposed DL models with the ReLU activation function and z-score and L2 normalizations are presented in Table \ref{tab:12}.  

\begin{figure*}[ht]
\centering
\begin{subfigure}[b]{0.38\textwidth}
    \centering
\includegraphics[width=\textwidth]{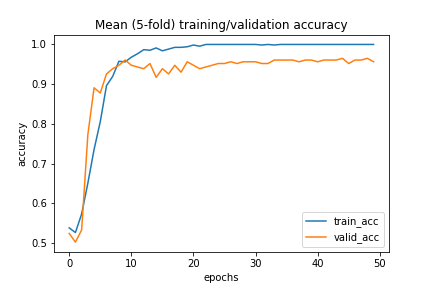}
\end{subfigure}
\hspace{7pt}
\begin{subfigure}[b]{0.38\textwidth}
    \centering
    \includegraphics[width=\textwidth]{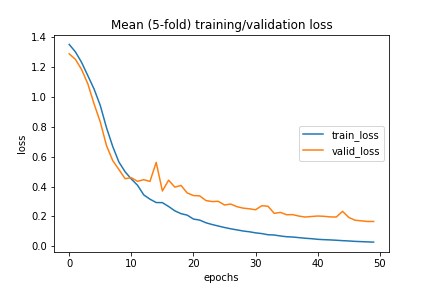}
\end{subfigure}
\caption{Learning curves of CNN-LSTM method with ReLU activation function and z-score normalization}
\label{fig:nz}

\end{figure*}

\begin{figure*}[h]
\centering
\begin{subfigure}[b]{0.38\textwidth}
    \centering
\includegraphics[width=\textwidth]{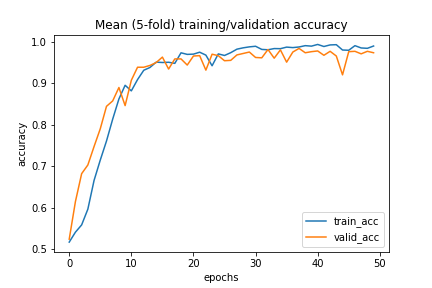}
\end{subfigure}
\hspace{7pt}
\begin{subfigure}[b]{0.38\textwidth}
    \centering
    \includegraphics[width=\textwidth]{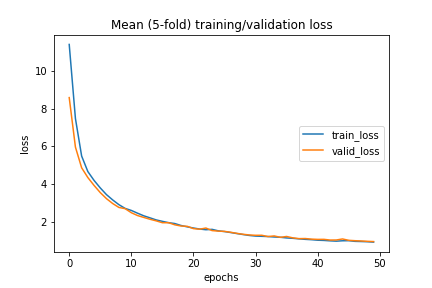}
\end{subfigure}
\caption{Learning curves of CNN-LSTM method with ReLU activation function and z-score + L2 normalization}
\label{fig:nzl}

\end{figure*}

According to Table \ref{tab:12}, it can be seen that compared to all classification methods with different activation functions, the second proposed CNN-LSTM model with ReLU activation function and combined normalization technique of z-score and L2 could lead to the maximum accuracy. In the following, the ROC diagrams for the DL models with ReLU activation functions and z-score and z-score + L2 normalization methods are drawn in Figure \ref{fig:6a} and \ref{fig:6b}. Furthermore, learning curves of the CNN-LSTM method with ReLU activation and z score normalization and also with z-score + L2 normalization are shown in Figures \ref{fig:nz}, \ref{fig:nzl} respectively.

\begin{figure}[ht]
\centering
\includegraphics[width=3.4in]{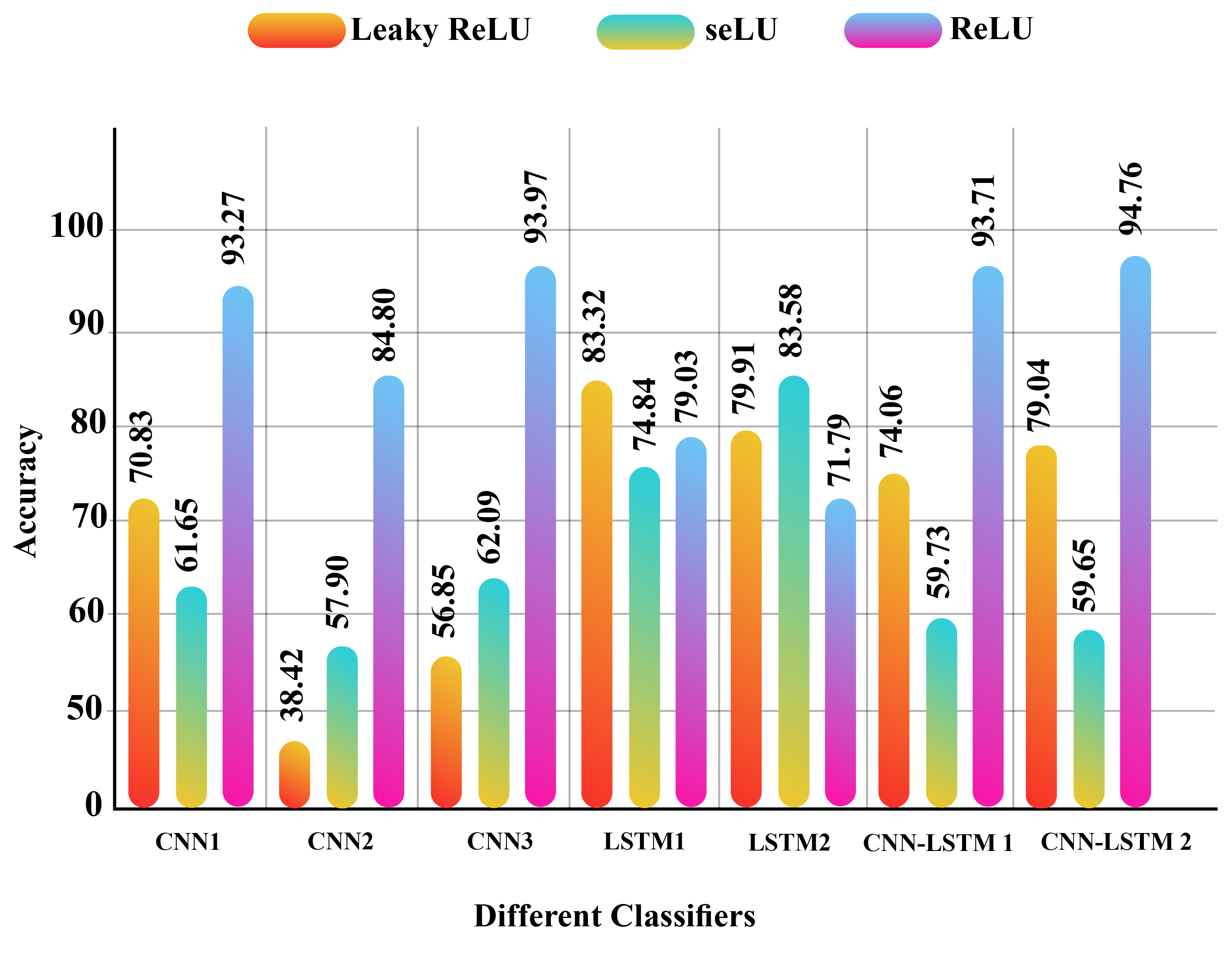}
\caption{Results for different proposed DL methods with different activation functions and z-score normalization.}
\label{fig:7}
\end{figure}

\begin{figure}[ht]
\centering
\includegraphics[width=3.4in]{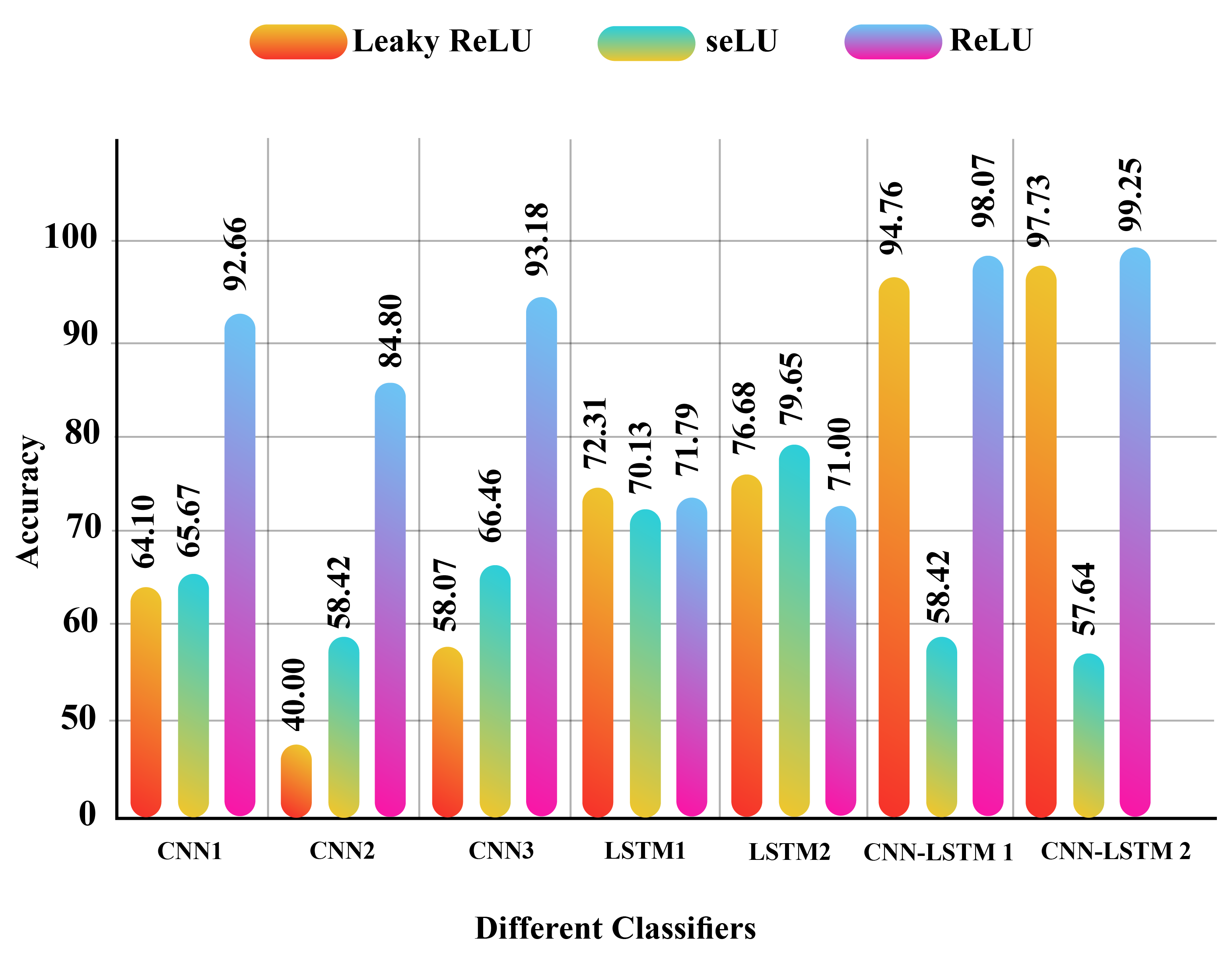}
\caption{Results for different proposed DL methods with different activation functions and z-score with L2 normalization.}
\label{fig:8}
\end{figure}

The simulation results of the proposed models for SZ diagnosis via EEG signals were investigated in this section. Compared to all DL and conventional ML methods, the CNN-LSTM models with 13 layers have higher accuracy and efficiency among the proposed methods. Selecting the number of layers in this model and the type of the activation functions are presented in this research for the first time, which is the article's novelty. Besides, simultaneously using z-score and L2 normalizations along with the proposed CNN-LSTM model is another novelty of this article. Figure \ref{fig:7} shows the DL models with different activation function and z-score normalization. Also, Figure \ref{fig:8} displayed the DL architectures with different activation functions and z-score and L2 normalization. According to Figures \ref{fig:7} and \ref{fig:8}, the second version of CNN-LSTM with z-score and L2 normalization has the best performance compared to other methods. 

\section{Limitation of Study}
\begin{table*}[ht]
\centering
\caption{Performance criteria of the proposed DL methods with ReLU activation function.}
\resizebox{\linewidth}{!}{
\begin{tabular}{|c|c|c|c|c|c|c|c|c|}
\hline
\multicolumn{1}{|c|}{\multirow{2}{*}{Methods}} & \multicolumn{4}{c|}{ReLU + Z-Score} & \multicolumn{4}{c|}{ReLU + Z-Score + L2} \\
\cline{2-9}
\multicolumn{1}{|c|}{} & \multicolumn{1}{c|}{Acc} & \multicolumn{1}{c|}{Prec} & \multicolumn{1}{c|}{Rec} & \multicolumn{1}{c|}{AUC} & \multicolumn{1}{c|}{Acc} & \multicolumn{1}{c|}{Prec} & \multicolumn{1}{c|}{Rec} & \multicolumn{1}{c|}{AUC} \\
\hline
CNN-1      & 93.27 ± 1.31          & 90.15 ± 4.60  & 93.18 ± 5.18  & 97.80 ± 0.35  & 92.66 ± 1.39          & 92.01 ± 2.57 & 88.86 ± 6.15 & 97.40 ± 0.60  \\ \hline
CNN-2      & 84.80 ± 11.7          & 65.18 ± 32.79 & 78.63 ± 39.34 & 88.80 ± 19.40 & 84.80 ± 11.7          & 89.25 ± 2.55 & 85.84 ± 9.18 & 88.63 ± 8.71  \\ \hline
CNN-3      & 93.97 ± 2.33          & 89.16 ± 5.34  & 96.59 ± 2.87  & 97.74 ± 0.85  & 93.18 ± 1.25          & 89.33 ± 5.17 & 94.09 ± 4.21 & 98.04 ± 0.23  \\ \hline
LSTM-1     & 79.03 ± 3.92          & 69.71 ± 6.01  & 82.95 ± 4.711 & 87.76 ± 3.26  & 71.79 ± 7.83          & 67.12 ± 10.3 & 57.72 ± 28.8 & 73.71 ± 11.48 \\ \hline
LSTM-2     & 71.79 ± 8.72    & 50.58 ± 26.85 & 70.45 ± 35.26 & 77.31 ± 14.52 & 71.0 ± 12.16          & 69.48 ± 14.5 & 68.18 ± 31.3 & 76.37 ± 12.46 \\ \hline
CNN-LSTM 1 & 93.71 ± 0.71          & 89.09 ± 2.505 & 95.45 ± 1.901 & 96.37 ± 0.62  & 98.07 ± 1.47          & 96.01 ± 3.91 & 99.31 ± 0.55 & 99.88 ± 0.11  \\ \hline
CNN-LSTM 2 & 94.76 ± 1.23          & 90.79 ± 1.914 & 96.14 ± 1.541 & 97.29 ± 0.50  & \textbf{99.25 ± 0.25} & 98.33 ± 3.33 & 98.86 ± 1.24 & 99.73 ± 0.35  \\ \hline

\end{tabular}}
\label{tab:12}
\end{table*}


The limitations of the study are investigated in this section. The available EEG datasets for SZ diagnosis consist of a limited number of cases which has made access to the tools of SZ diagnosis via EEG signals and DL models challenging. The dataset in this research was not used to determine the disorder's severity but to diagnose the disorder. This dataset is unsuitable for prognosis or early diagnosis, and other appropriate datasets must be gathered for these purposes. Another limitation of this study is that the classifiers are not separately designed and compared for different age and gender groups, and other suitable datasets must be gathered for this purpose. Classifiers are of two-class type and can become multi-class by adding the classes of brain disorders with similar symptoms to SZ.

\section{Conclusion, Discussion, and Future Works}

\begin{table*}[!ht]
\centering
\caption{The proposed method compared with related works in diagnosis of schizophrenia.}
\resizebox{\linewidth}{!}{
    \begin{tabular}{lcccccc}
        \hline
Work & Dataset  & Number of Cases & Preprocessing    & Feature Extraction and Selection   & Classifer & Accuracy \\ 
\hline
\cite{a27}        & Kaggle         & SZ:49, HC:32    & EMD       & Statistical features +KW Test    & EBT       & 89.59  \\
\cite{a28}        & Clinical       & SZ:14, HC:14    & Filtering   & Non- Linear Features + T-Test & SVM-RBF   & 92.90   \\
\cite{a29}        & Clinical       & SZ:11, HC:9     & Filtering    & ERP features   & LDA       & 71.00    \\
\cite{a30}        & Clinical       & SZ:14, HC:14    & ICA   & Isomap + Opt. Methods  & Adaboost  & 98.77   \\
\cite{a31}        & Clinical       & SZ:23, HC:23    & NA    & \makecell{Statistical Features of\\SSVEPs + Fisher Score} & KNN   & 91.30  \\
\cite{a32}        & Clinical       & SZ:19, HC:23    & Filtering  & SPN Features    & SVM       & 90.48  \\
\cite{a33}        & Clinical       & SZ:5, HC:5      & NA   & Different Features    & LR        & NA    \\
\cite{a34}        & Clinical       & Different Cases & Interpolation Algorithms    & Microstate Features  & RF  & NA    \\
\cite{a35}        & Clinical       & SZ:34, HC:34    & Filtering     & Different features + Fisher Score     & SVM     & 88.24   \\
\cite{a36}        & Public Dataset & SZ:14, HC:14    & Filtering  & ResNet-18   & SVM       & 98.60   \\
\cite{a37}        & Clinical       & Different Cases & NA & CNN+LSTM  & Sigmoid   & 72.54  \\
\cite{a38}        & Clinical       & SZ:54, HC:55    & Filtering   & CNN-LSTM    & Softmax   & 99.22  \\
\cite{a39}        & Public Dataset & SZ:45, HC:39    & NA   & MDC-CNN  & Softmax   & 93.06   \\
\cite{a40}        & Clinical       & SZ:40, HC:40    & Ocular Correction, Filtering  & CNNV  & RF   & 99.20 \\
\cite{a41}        & Clinical       & SZ:14, HC:14    & Z-score Normalization  & CNN  & Softmax   & 89.59 \\
\cite{a42}        & NNCI           & SZ:45, HC:39    & \makecell{Pearson Correlation\\Coefficient (PCC)} & CNN  & Softmax   & 90.00  \\
\cite{a43}        & Clinical       & SZ:21, HC:24    & Filtering  & CNN-LSTM   & Sigmoid   & 99.10         \\
\cite{a44}        & NNCI           & SZ:45, HC:39    & Filtering   & CNN-LSTM   & Sigmoid   & 98.56   \\
\cite{a45}        & NNCI           & SZ:45, HC:39    & NA     & DBN   & Softmax   & 95.00   \\ 
\makecell{Proposed\\Method} & Public Dataset & SZ:14, HC:14    & Filtering, Normalization    & 1D CNN-LSTM     & Sigmoid   & 99.25 \\
\hline
    \end{tabular}}
    \label{tab:13}
\end{table*}

SZ is a mental disorder that negatively affects brain function, causing various problems for the patient. Different screening methods have been introduced for SZ mental disorder diagnosis, among which the EEG functional imaging modality has captured the interest of neurologists and specialist physicians. SZ diagnosis via EEG signals has always been challenging. In recent years, various investigations into using AI techniques for SZ diagnosis and interpretation via EEG signals have been conducted to tackle this challenge. These methods are proposed to help physicians and neurologists with quick and accurate diagnosis of SZ disorder via EEG signals. 


Various AI approaches are presented for diagnosis of SZ mental disorder via EEG signals. These approaches include using different conventional ML techniques \cite{a67} and also DL models \cite{a68,a69,ho1,ho2}. The AI models for SZ diagnosis via EEG signals consist of the following steps: dataset selection, preprocessing, feature extraction and selection, and classification. 

In this study, the dataset consisted of EEG data of 14 normal individuals and patients with SZ \cite{a46}. The EEG signals of this dataset are of a 10-channel type and have a sampling frequency of 250 Hz \cite{a46}. In the preprocessing step, first, the EEG signals were divided into 25-second frames. Afterward, z-score and z-score-L2 were used for the normalization of EEG signals. In this section, each frame of EEG signals had a dimension of 19×6250. It should be noted that the preprocessing of EEG signals for the DL models included two z-score and z-score-L2 normalization techniques.

Different conventional ML-based classification algorithms were used in for SZ diagnosis via EEG signals. In this section, the normalized EEG signals were considered as features to be applied in classification algorithms. The employed classification algorithms included the following methods: SVM \cite{a47}, KNN \cite{a48}, DT \cite{a49}, naïve Bayes \cite{a50}, RF \cite{a51}, ERT \cite{a52}, and bagging \cite{a53}. The algorithms were evaluated using a 5-fold cross validation strategy, and the bagging classification via EEG signals normalized using z-score could obtain an accuracy of 81.22 ± 1.74, which is the highest accuracy compared to other classification methods.

In the following, different DL methods of SZ diagnosis via EEG signals were employed. The proposed DL methods in this section included three 1D-CNN architectures, two LSTM models, and ultimately two 1D-CNN-LSTM networks. Different activation functions, including Leaky ReLU, seLU, and ReLU were used to implement the proposed DL models. Besides, in all models, the sigmoid activation function was used for classification. The results of DL models for different normalization methods and activation functions were evaluated again using a 5-fold cross validation strategy, and the results were indicated in Tables \ref{tab:ten} to \ref{tab:12}. Among the proposed DL models, the 1D-CNN-LSTM architecture consisting of 13 layers with the ReLU activation function and Z-Score + L2 normalization could obtain an accuracy of 99.25 ± 0.25. This model is presented for the first time in this research, as this article's novelty. The comparison between the proposed 1D-CNN-LSTM model with the proposed models of the previous studies conducted on SZ diagnosis via EEG signals is indicated in Table \ref{tab:13}.


As shown in Table \ref{tab:13}, the model proposed in this research could obtain higher accuracy compared to a vast majority of conducted studies. The proposed model can be implemented on special software and hardware platforms for quick SZ diagnosis via EEG signals and may be employed as an assistant diagnosis method in hospitals. 

In the following, some future investigations into SZ diagnosis via EEG signals are presented. The CNN-AE models can be employed for SZ diagnosis via EEG signals as the first future work. Several researchers indicate that CNN-AE models are highly efficient in neural disorders via EEG signals \cite{a25}. As mentioned in the section of limitation of the study, the dataset used in this study is for SZ disorder diagnosis. However, providing EEG datasets for SZ disorder diagnosis can be of paramount importance for future investigations. One of the future works is to provide classification models based on DL for different age and gender groups, which requires researchers to have access to relevant data.

Another future work is using a combination of conventional ML and DL models for SZ diagnosis such that different non-linear features are extracted from EEG signals first. Afterward, the features are extracted from raw EEG signals by DL models. Ultimately, handcrafted and DL features are combined, and the classification is carried out. Graph models based on DL are one of the new fields in diagnosing brain disorders. Accordingly, in future works, using graph models based on DL can be suitable for SZ diagnosis via EEG signals \cite{a66}.

\section*{Acknowledgment}

This work was partly supported by the MINECO/ FEDER under the RTI2018-098913-B100 CV20-45250 and A-TIC- 080-UGR18 projects.


%





\ifCLASSOPTIONcaptionsoff
  \newpage
\fi





\bibliographystyle{IEEEtran}
\bibliography{main}

\vfill


\end{document}